\documentclass[conference]{IEEEtran}
\IEEEoverridecommandlockouts

\usepackage{amsmath,amsfonts,amssymb,amsopn}
\usepackage{graphicx}
\usepackage{xcolor}
\usepackage{cite}
\usepackage{xspace}
\usepackage{subfig} 
\usepackage{mathtools} 
\usepackage{url}
\usepackage{algorithm}
\usepackage{algpseudocode}
\usepackage{ifpdf}

\usepackage{float}
\floatname{algorithm}{Procedure}

\ifCLASSOPTIONonecolumn

\renewcommand{\baselinestretch}{2}
\usepackage[margin=25mm]{geometry}
\else

\fi

\newcommand{\bv}[1]{\boldsymbol{#1}}
\newcommand{\dfn}{\triangleq}
\newcommand{\untsph}{\mathbb{S}^{2}} 

\newcommand{\shc}[3]{({#1})_{#2}^{#3}}
\newcommand{\lsph}{L^2(\untsph)}
\newcommand{\conj}[1]{\overline{#1}} 
\newcommand{\unit}[1]{\bv{\hat{#1}}}
\newcommand{\lsphL}[1]{\mathcal{H}_{#1}}

\newcommand{\intsph}{\int_{\untsph}}

\newcommand{\figref}[1]{Fig.\,\ref{#1}}

\DeclarePairedDelimiterX\innerp[2]{\langle}{\rangle}{#1,#2}

\graphicspath{{figs/},{pdfs/},{pdf/},{Figures/},{Authors/}}

\begin{document}
\title{Iterative Residual Fitting for Spherical Harmonic Transform of Band-Limited Signals on the Sphere: Generalization and Analysis}

\author{
\IEEEauthorblockN{Usama Elahi\IEEEauthorrefmark{1},
Zubair Khalid\IEEEauthorrefmark{2},
Rodney A. Kennedy\IEEEauthorrefmark{1} and Jason D. McEwen\IEEEauthorrefmark{3}}

\IEEEauthorblockA{\IEEEauthorrefmark{2}Research School of Engineering, The Australian National University, Canberra, ACT 2601, Australia }
\IEEEauthorblockA{\IEEEauthorrefmark{1}School of Science and Engineering, Lahore University of Management Sciences, Lahore 54792, Pakistan}
\IEEEauthorblockA{\IEEEauthorrefmark{3}Mullard Space Science Laboratory, University College London, Surrey RH5 6NT, UK
\thanks{Usama Elahi, Zubair Khalid and Rodney A. Kennedy are supported by Australian Research Council's Discovery Projects funding scheme (project no. DP150101011). Jason D. McEwen is partially supported by the Engineering and Physical Sciences Research Council (grant number EP/M011852/1).}
}

\textit{usama.elahi@anu.edu.au, zubair.khalid@lums.edu.pk, rodney.kennedy@anu.edu.au, jason.mcewen@ucl.ac.uk}}



\maketitle

\begin{abstract}

We present the generalized iterative residual fitting~(IRF) for the computation of the spherical harmonic transform~(SHT) of band-limited signals on the sphere. The proposed method is based on the partitioning of the subspace of band-limited signals into orthogonal subspaces. There exist sampling schemes on the sphere which support accurate computation of SHT. However, there are applications where samples~(or measurements) are not taken over the predefined grid due to nature of the signal and/or acquisition set-up. To support such applications, the proposed IRF method enables accurate computation of SHTs of signals with randomly distributed sufficient number of samples. In order to improve the accuracy of the computation of the SHT, we also present the so-called multi-pass IRF which adds multiple iterative passes to the IRF.  We analyse the multi-pass IRF for different sampling schemes and for different size partitions. Furthermore, we conduct numerical experiments to illustrate that the multi-pass IRF allows sufficiently accurate computation of SHTs.
	
\end{abstract}

\begin{IEEEkeywords}
     Spherical harmonics, basis functions, spherical harmonic transform, residual fitting, band-limited signals, 2-sphere~(unit sphere).
\end{IEEEkeywords}

\section{Introduction}

Signals are defined on the sphere in a variety of fields including geodesy~\cite{Amirbekyan:2008}, computer graphics~\cite{Ng:2004}, cosmology~\cite{Yadav}, astrophysics~\cite{Jarosik:2010}, medical imaging~\cite{Chung:2007}, acoustics~\cite{Zhang:2012} and wireless communication~\cite{Yibe:2015} to name a few. Spherical harmonic~(SH) functions~\cite{Kennedy-book:2013} are a natural choice of basis functions for representing the signal on the sphere in all these applications. Analysis on the sphere is done in both spatial~(spherical) and spectral (spherical harmonic) domains.  The transformation between the two domains is enabled by the well known spherical harmonic transform~(SHT)~\cite{Kennedy-book:2013,Sakurai:1994}. For harmonic analysis and signal representation~(reconstruction), the ability to accurately compute the SHT of a signal from its samples taken over the sphere is of great importance.

Sampling schemes have been devised in the literature for the accurate and efficient computation of SHTs~\cite{Equi:2011,khalid:2014}. However, the samples may not be available, in practice~(e.g.,~\cite{Sneeuw:1994,Chung:2007}), over the grid defined by these sampling schemes. To support the computation of SHTs in applications where samples or data-sets are not available on the pre-defined grid, least squares fitting~(LSF) methods have been investigated for efficient computation of the SHTs~\cite{Sneeuw:1994,Healy:2003,Blais:2006,Ivanov:2014,Keiner:2007,Kunis:2003,Kunis:2007}. LSF methods formulate a large linear system of basis functions and then attempt to solve it efficiently. However, due to memory overflow, it is not suitable for systems with large band-limits, $L>1024$~\cite{Shen:2006,Black,Barret}. To solve this problem,  an iterative residual fitting~(IRF) method has been proposed in \cite{Shen:2006} as an extension of LSF and incorporates a divide and conquer technique for the computation of SHTs. The basic idea of IRF is to divide the subspace spanned by all spherical harmonics into smaller partitions and then perform least squares on each partition iteratively. Although IRF is fast, it creates a less accurate reconstruction~\cite{Shen:2006} as the size of the harmonic basis increases for large band-limits. To improve the reconstruction accuracy, a multi-pass IRF approach is used which includes multiple passes for fitting. This is same as IRF but it involves multiple IRF operations rather than one. A variant of this scheme is presented in~\cite{Shen:2006}, where reconstruction for 3D surfaces is carried out by taking large number of samples.

In this paper,  we present an IRF method for the computation of the SHT of a band-limited signal in a general setting that partitions the subspace of band-limited signals into orthogonal subspaces, where each orthogonal subspace can be spanned by different numbers of basis functions.  We also formulate multi-pass IRF to improve the accuracy of computation of the SHT. We analyze multipass IRF for different choices of partitioning of the subspace and sampling schemes~\cite{Equi:2011,khalid:2014,Healpix:2005,Shen:2006} and show that the computation of the SHT converges in all cases. We also show that the convergence is fast for the partition choice considered in this work.

The remainder of this paper is organized as follows. Section 2 provides the necessary mathematical background and notation required to understand the work. IRF and multi-pass IRF methods are formulated in section 3. In section 4, we carry out accuracy analysis of the proposed IRF method for different partition choices and sampling schemes. Finally, concluding remarks are presented in section 5.

\section{Mathematical Background}\label{sec:models}
\subsection{Signals on the Sphere}
\label{sec:models:signals:sphere}
A point $\unit{v}=\mathbf{\unit{v}}(\theta,\phi)$ on the unit sphere  $\mathbb{S}^2 \triangleq\{\mathbf{\unit{v}} \in \mathbb{R}^3 \colon |
\mathbf{\unit{v}}| = 1  \}$, is parameterized by $[\sin\theta \cos\phi,\cos\theta \cos\phi,\cos\theta]^T\in \mathbb{S}^2 \subset \mathbb{R}^3$, where $(.)^T$ represents the transpose, $\theta \in [0, \pi]$ represents the co-latitude and $\phi\in [0, 2\pi)$ denotes the longitude. The space of square integrable complex functions of the form $g(\theta,\phi)$, defined on  the unit sphere, form a complex separable Hilbert space, denoted by $\lsph$, with inner product defined as by~\cite{Kennedy-book:2013}
\begin{align}\label{eqn:innprd}
	\langle g, h \rangle \triangleq  \int_{\mathbb{S}^2}
	g(\theta,\phi) \overline {h(\theta,\phi)}
	\,\sin\theta\,d\theta\,d\phi, \quad g,h\in\lsph,
\end{align}
where $\overline{(\cdot)}$ represents the complex conjugate operation. The functions with finite induced norm $\|g\| \triangleq\langle g,g \rangle^{1/2}$ are referred to as signals on the sphere.

	
\subsection{Spherical Harmonics}
\label{sec:harmonic_standard_expansion}
Spherical harmonic~(SH) functions, denoted by $Y_\ell^m(\theta,\phi)$ for integer degree $\ell \ge 0$ and integer order $ |m| \le \ell$, serve as complete basis for $\lsph$~\cite{Kennedy-book:2013}. Due to the completeness of the SH functions, any function $g$ on the sphere can be expanded as
\begin{equation}
	\label{Eq:f_expansion}
	g(\theta,\phi)=\sum_{{\ell}=0}^{\infty}\sum_{m=-{\ell}}^{\ell} \shc{g}{\ell}{m}  Y_{\ell}^m(\theta,\phi),
\end{equation}
where $\shc{g}{\ell}{m}$ are the SH coefficients of degree $\ell$ and order $m$ and form the spectral domain representation of the signal $g$, given by the spherical harmonic transform~(SHT) defined as
\begin{equation}\label{Eq:fcoeff}
	\shc{g}{\ell}{m}\dfn\innerp[\big]{f}{Y_{{\ell}}^m} =
	\intsph f(\theta,\phi)\conj {Y_{\ell}^m(\theta,\phi)} \,\sin\theta\,d\theta\,d\phi.
\end{equation}
The signal $g$ is band-limited at degree $L$ if $\shc{g}{\ell}{m} = 0$ for all $\ell \ge L$, $|m| \leq {\ell}$. A set of band-limited signals forms an $L^2$ dimensional subspace of $\lsph$, denoted by $\lsphL{L}$.

\section{Generalized Iterative Residual Fitting}
Here we present the generalization of the IRF method~\cite{Chung:2007,Shen:2006} for the computation of the SHT of the band-limited signal $g\in\lsphL{L}$ from its samples.

\begin{figure}[t]
\centering
\includegraphics[scale=0.32]{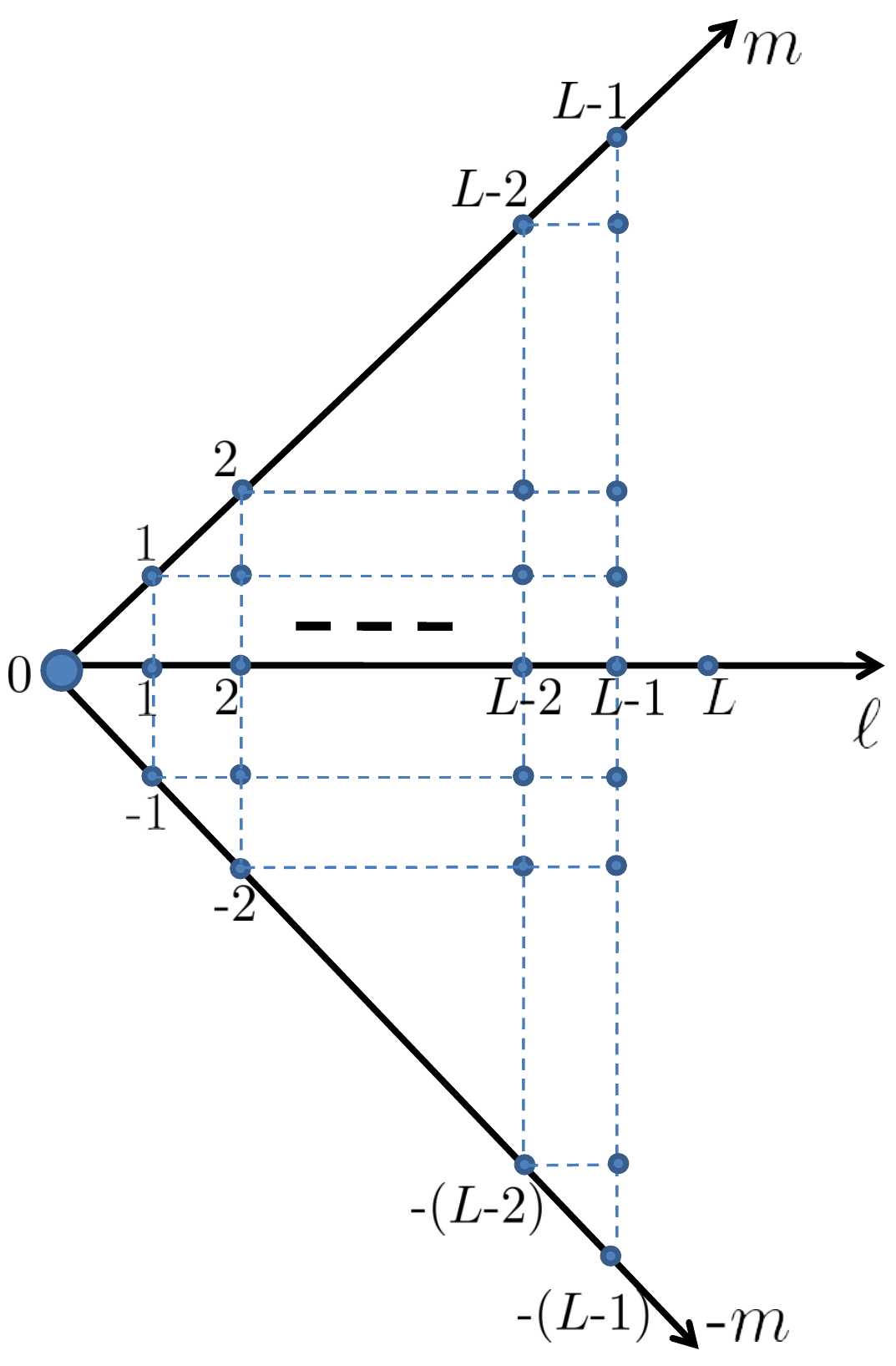}
\caption{Spherical harmonic domain representation of a band-limited signal in $\lsphL{L}$. }
\label{fig:spectral_domain_representation}
\end{figure}

\subsection{Iterative Residual Fitting~(IRF) -- Formulation}
The IRF method is based on the idea to partition the subspace $\lsphL{L}$ into smaller subspaces and carry out least-squares estimation on these partitions iteratively. In this way, a large linear problem is divided into manageable small subsets of linear problems. The subspace $\lsphL{L}$ has graphical representation of the form shown in  \figref{fig:spectral_domain_representation}, which also represents the SH~(spectral) domain formed by the SH coefficients of the band-limited signal in $\lsphL{L}$. We partition $\lsphL{L}$ into $K$ \emph{orthogonal} subspaces $\lsphL{L}^k,\, k=1,2,\cdots, K$, each of dimension $N_k$. We analyse different choices for partitioning later in the paper. We index the SH functions that span the subspace $\lsphL{L}^k$ as $Y_{k j},\,j=1,2,\cdots,N_k$. We also define $({g})_{k j} = \innerp[\big]{g}{Y_{k j}}$.

Given $M$ samples~(measurements) of the band-limited signal $g\in\lsphL{L}$, we wish to compute SH coefficients. By defining a vector
\begin{equation}\label{func}
	\mathbf{G} \dfn \big[g(\theta_1,\phi_1),\hdots,\,g(\theta_M,\phi_M)]^T,
\end{equation}
of $M$ measurements~(samples) of the signal $g\in\lsphL{L}$ on the sphere and the matrix $\mathbf{Y}_k$, with entries $\{ \mathbf{Y}_ {k} \}_{p,q} =  Y_{k q}(\theta_p,\phi_q)$, of size $M\times N_k$ containing SH functions that span the subspace $\lsphL{L}^k$ evaluated at $M$ sampling points, the vector $\mathbf{{g}_{k}}=[{g}_{k1},{g}_{k2},\cdots,{g}_{k N_k}]^T$ of SH coefficients can be \emph{iteratively} computed~(estimated) in the least-squares sense as
\begin{align}\label{Eq:IRF_ls}
\mathbf{\tilde{g}}_{k}=({\mathbf{Y}_{k}^H \mathbf{Y}_{k}})^{-1} \ \mathbf{Y}_{k}^H \ \mathbf{r}_{k},
\end{align}
where $(.)^H$ represents the Hermetian of a matrix and
\begin{align}\label{eq:res_IRF}
	\mathbf{r}_{k}=\mathbf{G}-  \sum_{k'=1}^{k-1}\mathbf{Y}_{k'} \    \mathbf{\tilde{g}}_{k'}, \quad \mathbf{r}_{0}  = \mathbf{G}
\end{align}
is the residual between the samples of the signal and the signal obtained by using the  coefficients $\mathbf{\tilde{g}}_{k'}$ for $k'=1,2,\hdots,k-1$ and the estimation of
coefficients is carried out iteratively for $k=1,2,\hdots, K$. We note that the computational complexity for \eqref{Eq:IRF_ls} for each $k$ would be of the order of $max(\mathbb{O}(MN_k^2),\mathbb{O}(N_k^3))=\mathbb{O}(MN_k^2)$. The computational complexity to compute \eqref{eq:res_IRF} is $\mathbb{O}(ML^2)$. We later analyse the estimation accuracy of the IRF method for different sampling schemes on the sphere and different partitions of the subspace $\lsphL{L}$ of band-limited signals. For a special case of partitioning the subspace $\lsphL{L}$ into $L$ subspaces $\lsphL{L}^k$ based on the degree of spherical harmonics $\ell=0,1,\hdots, L-1$, it has been shown that the iterative residual fitting allows sufficiently accurate estimation of SH coefficients~\cite{Shen:2006}.

The proposed IRF method enables accurate computation of the SHT of signals with a sufficient number of randomly distributed samples. The IRF algorithm finds significance use in applications where samples on the sphere are not taken over a predefined grid. For example, the samples are taken over the cortical surface in medical imaging~\cite{Chung:2007}, where IRF allows sufficient accurate parametric modeling of cortical surfaces.


\subsection{Multi-Pass IRF and Residual Formulation}
To improve the estimation accuracy, we employ the so-called multi-pass IRF~\cite{Shen:2006} which is based on the use of IRF method in an iterative manner. In multi-pass IRF, the IRF algorithm is run for a number of iterations, denoted by $i=1,2,\hdots$. To clarify the concept, we incorporate the iteration index $i$ in the formulation in \eqref{Eq:IRF_ls} and \eqref{eq:res_IRF} as
\begin{align}\label{Eq:IRF_ls_mp}
\mathbf{\tilde{g}}_{k}(i)=({\mathbf{Y}_{k}^H \mathbf{Y}_{k}})^{-1} \ \mathbf{Y}_{k}^H \ \mathbf{r}_{k}(i),
\end{align}
\begin{align}\label{eq:res_IRF_mp}
	\mathbf{r}_{k}(i)=\mathbf{G}-  \sum_{i'=1}^{i-1} \sum_{k'=1}^{k-1}\mathbf{Y}_{k'} \    \mathbf{\tilde{g}}_{k'}(i'), \nonumber
	 \\	 \mathbf{r}_{0}(i)  = \mathbf{r}_{K}{(i-1)},\,\,\mathbf{r}_{0}({1})  = \mathbf{G}.
\end{align}
After $i$-th iteration, $\mathbf{\tilde{g}}_{k}$ can be computed for each $k=1,2,\hdots,K$ as
\begin{align}
\mathbf{\tilde{g}}_{k}(i) = \sum_{i'=1}^{i} \mathbf{\tilde{g}}_{k}{(i')}.
\end{align}
By defining
\begin{align}
	\mathbf{A}_k \dfn {(\mathbf{Y}_{k}^H \ \mathbf{Y}_{k})}^{-1} \ \mathbf{Y}_k^H,\quad \mathbf{C}_{k} \dfn {\mathbf{Y}_{k} \ \mathbf{A}_{k}},
\end{align}
the residual after the $i$-th iteration is given by
\begin{equation}\label{eq:res}
r_{K}{(i)}=\left(\prod\limits_{k=1}^{K} (1 - \mathbf{C_k})\right)^i \mathbf{G}.
\end{equation}
In general, the residual in \eqref{eq:res} depends on the distribution of sampling points and nature of partitioning of $\lsphL{L}$. In the next section, we show that the residual converges to zero for a variety of sampling schemes and different partitions.

\begin{figure*}[!ht]
\centering
\hspace{-4mm}
\subfloat[]{
    \includegraphics[width=0.25\textwidth]{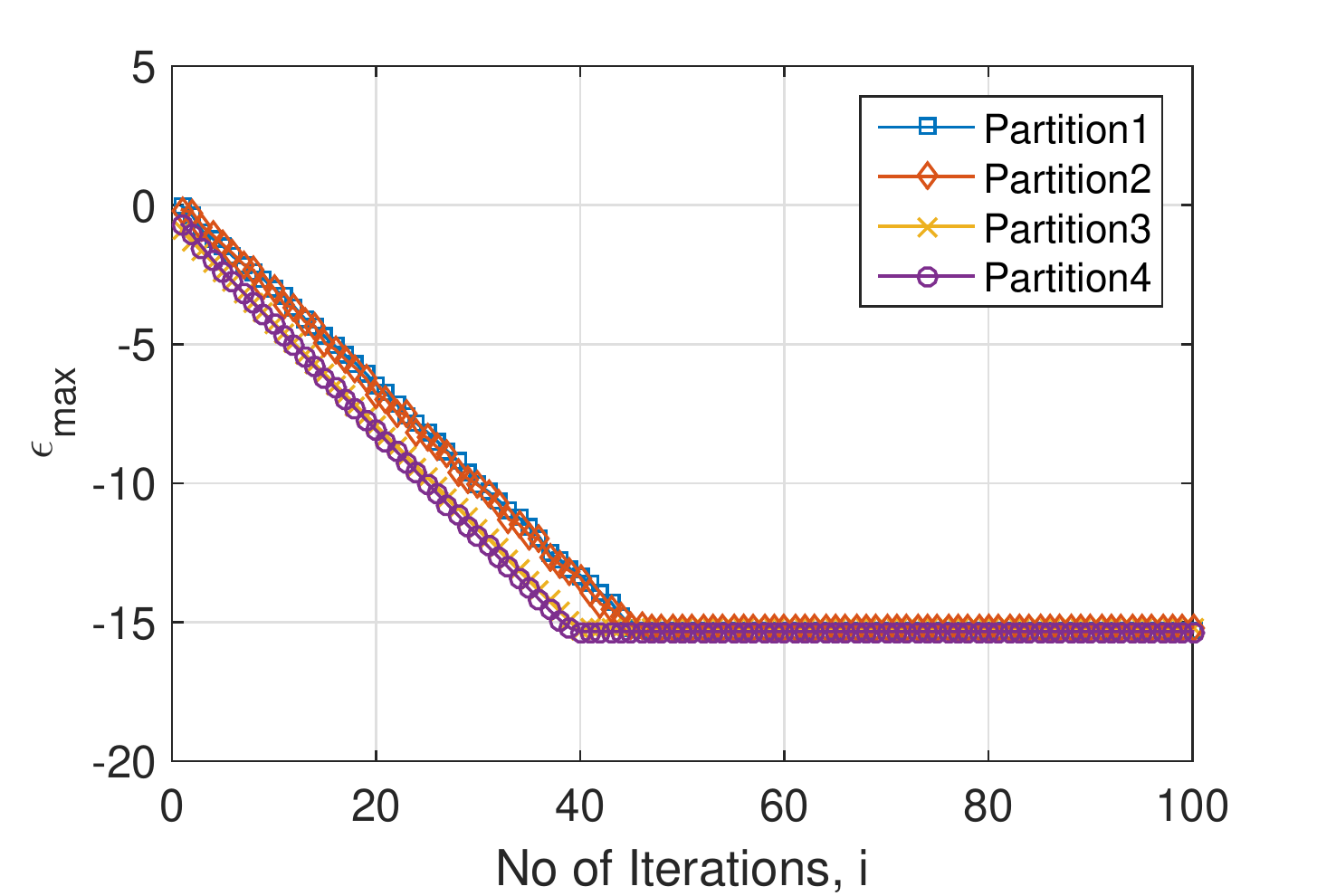}}\hfil
    \hspace{-1mm}
\subfloat[]{
    \includegraphics[width=0.25\textwidth]{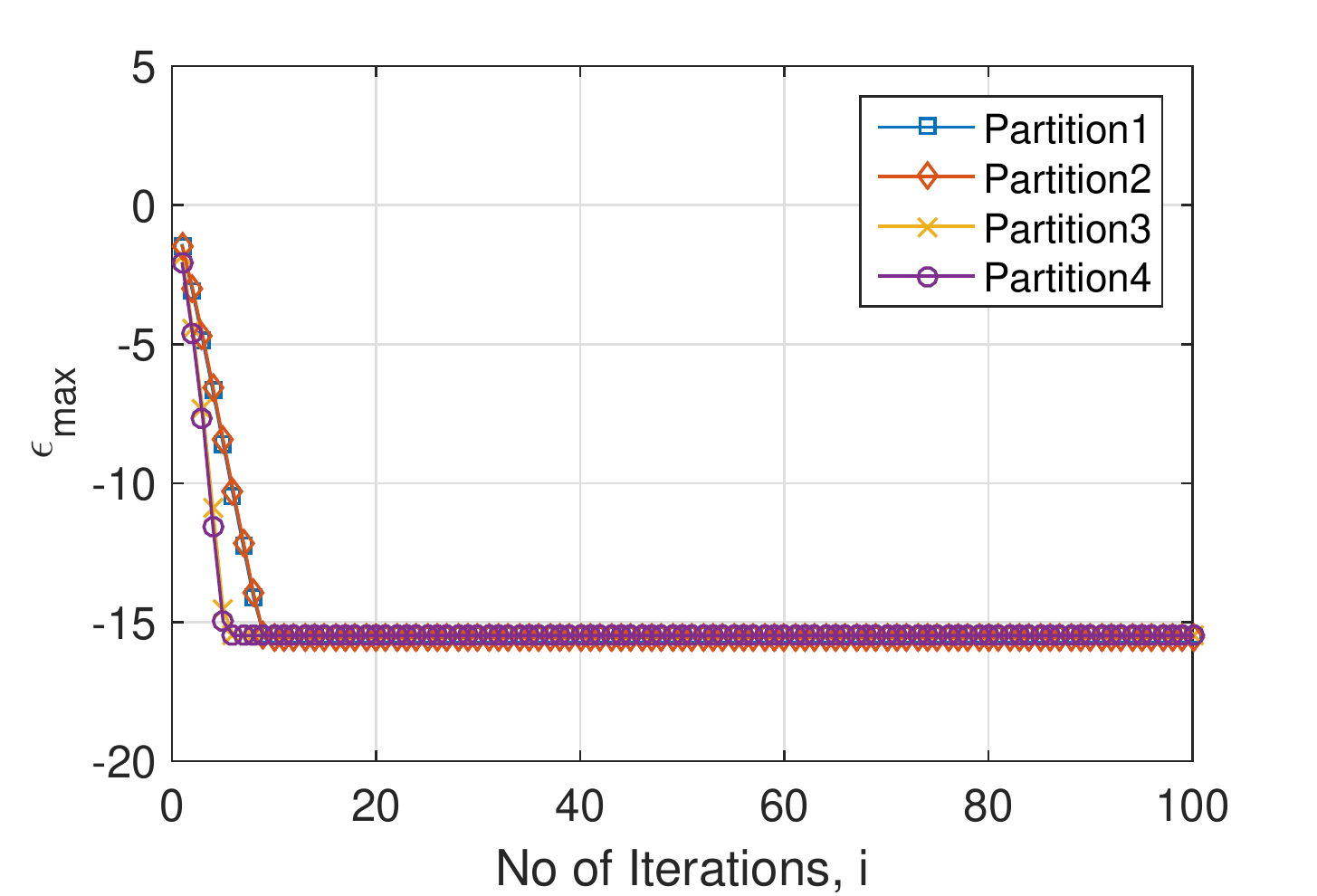}}\hfil
    \hspace{-1mm}
\subfloat[]{
    \includegraphics[width=0.25\textwidth]{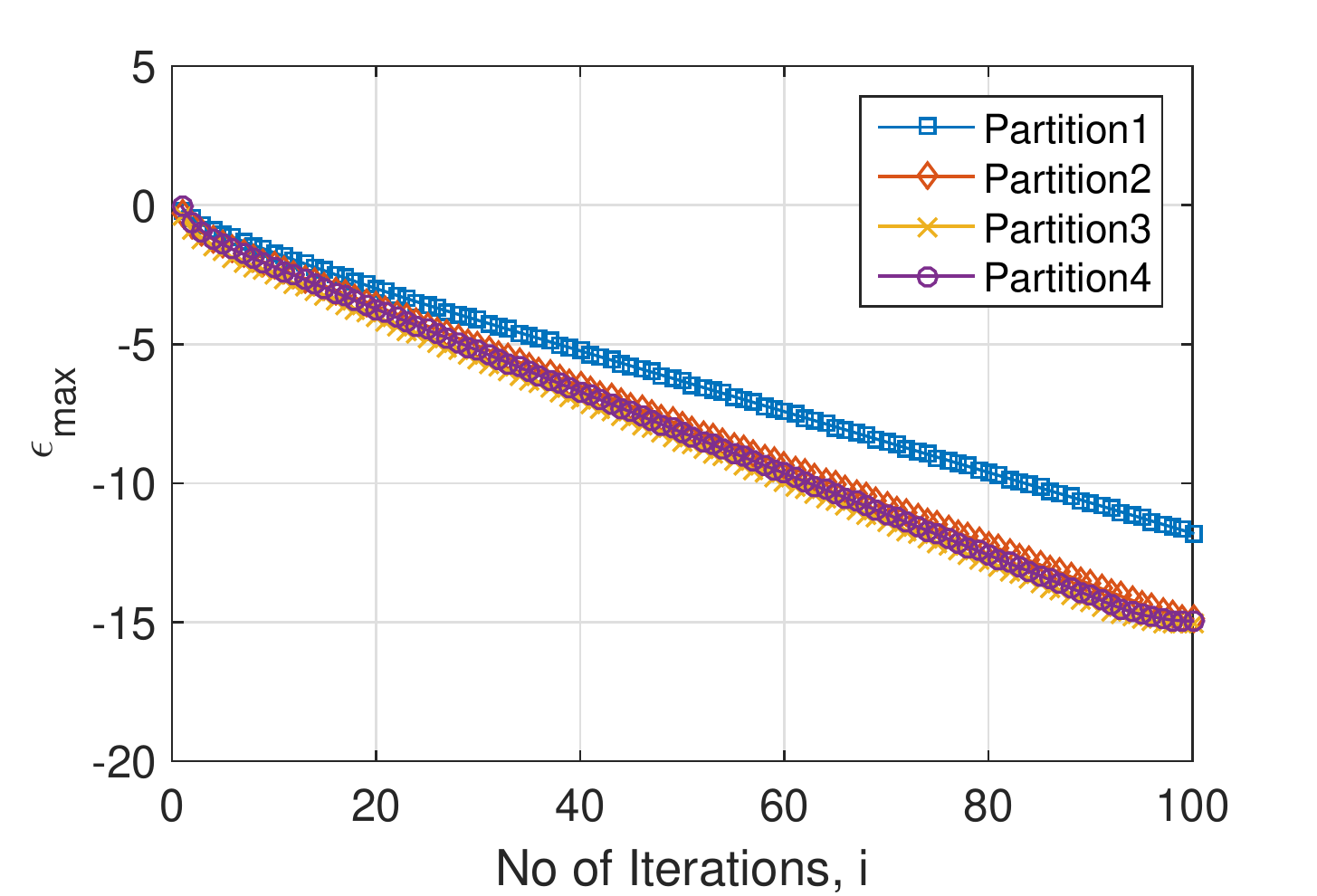}}\hfil
\subfloat[]{
   \includegraphics[width=0.25\textwidth]{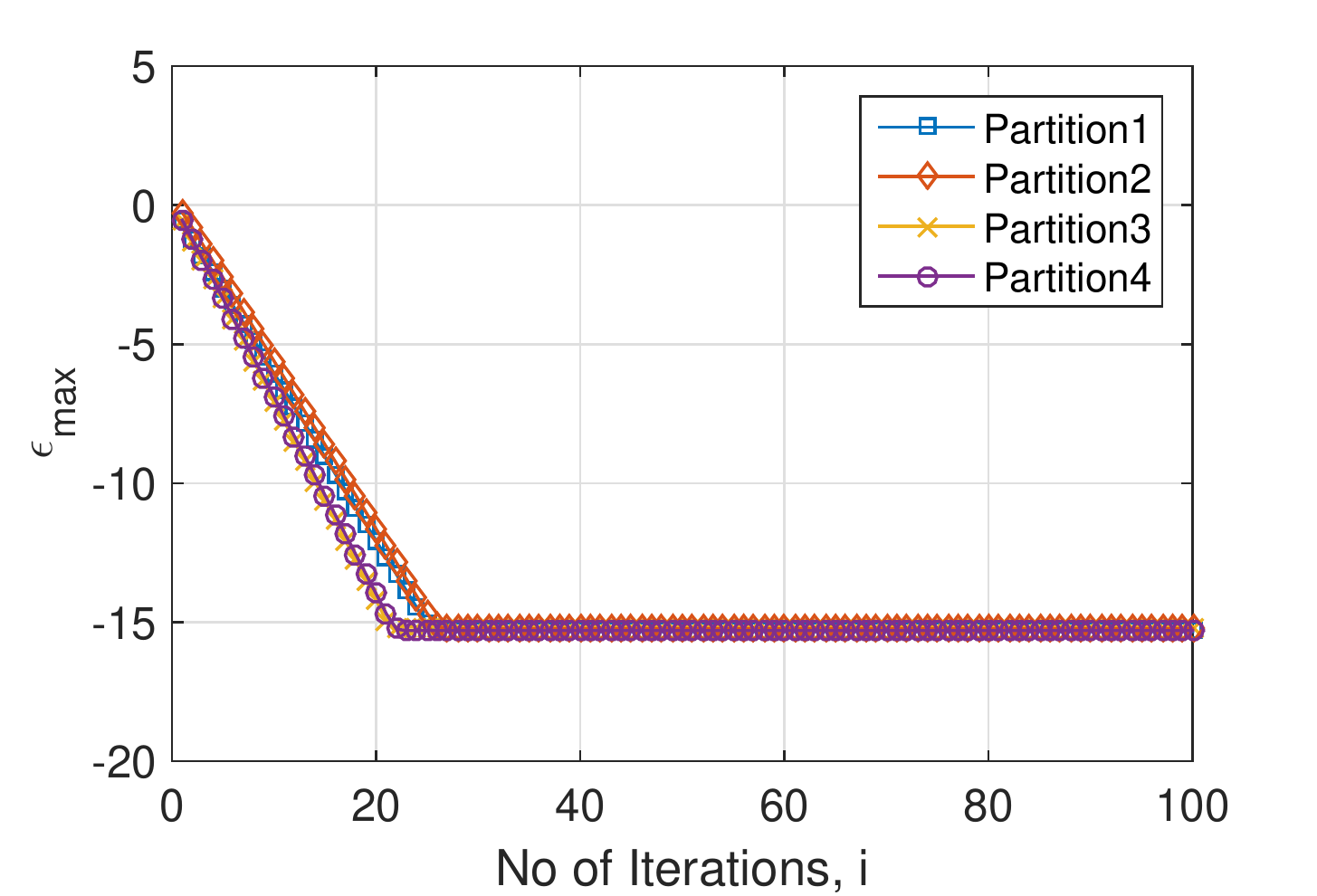}}\hfil
\subfloat[]{
\includegraphics[width=0.25\textwidth]{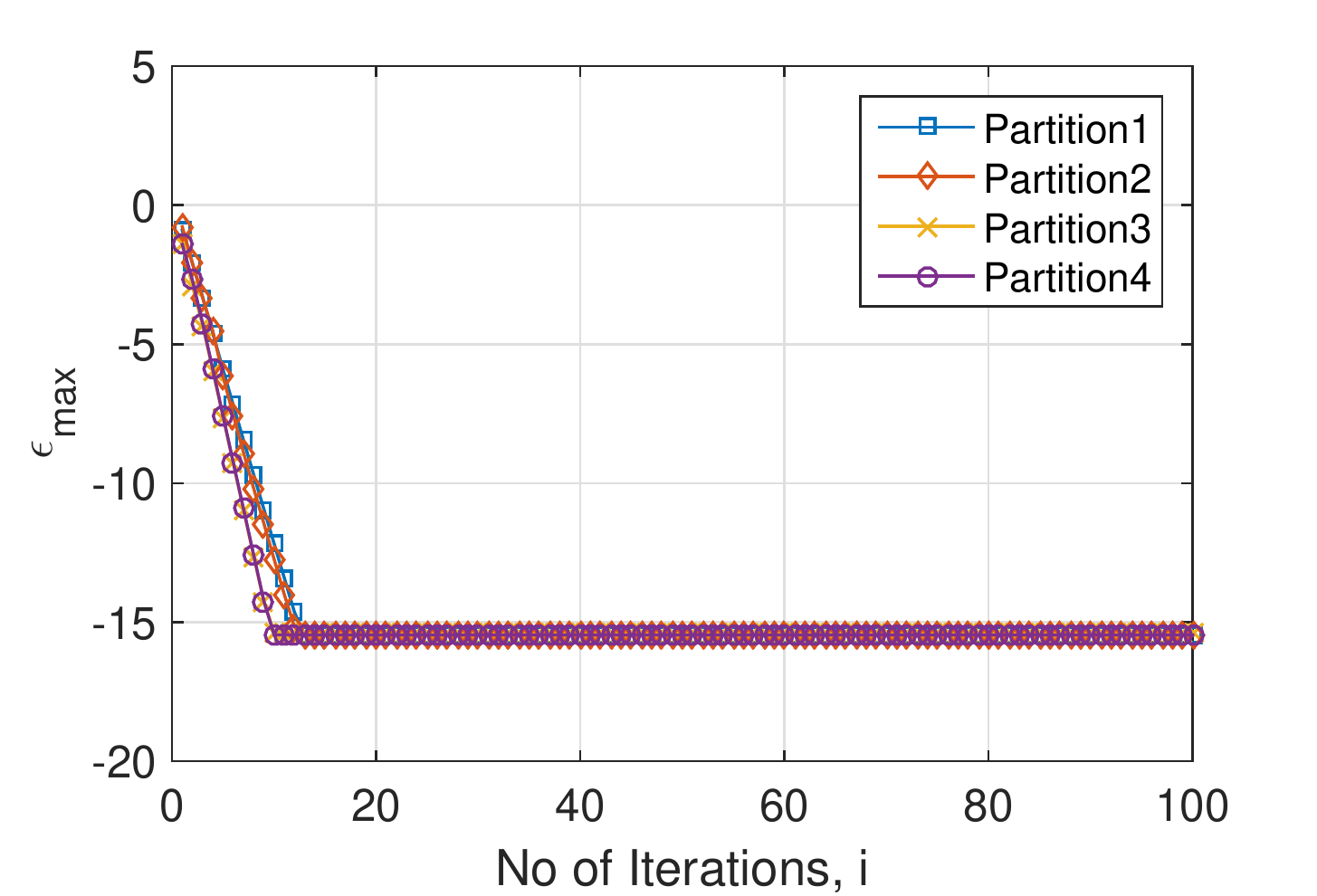}}\hfil
\subfloat[]{
\includegraphics[width=0.25\textwidth]{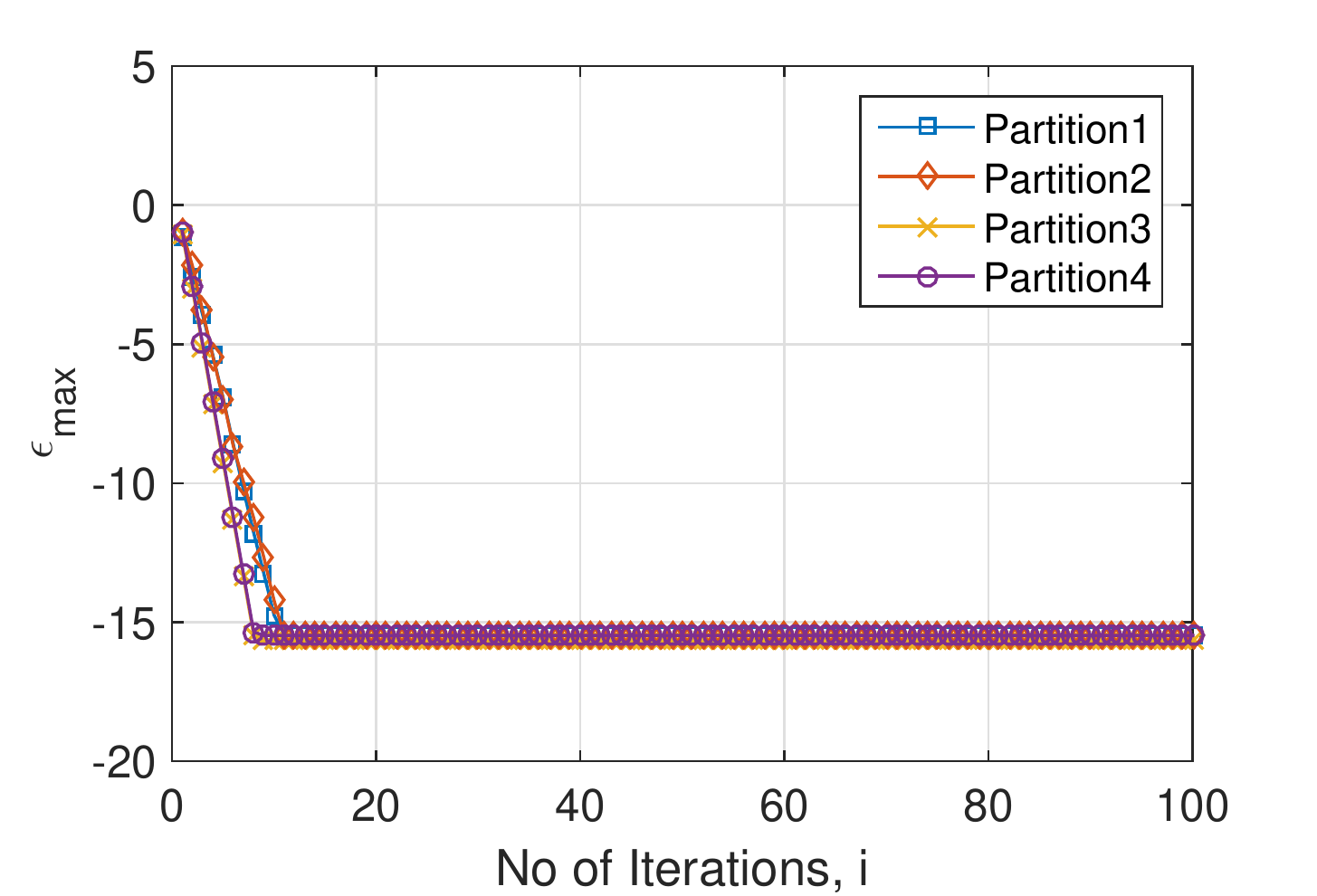}}\hfil
    \caption{Maximum reconstruction error $\epsilon_\textrm{max}$, given in
    \eqref{Eq:exp1:errors:max}, between the original and reconstructed SH coefficients of a band-limited signal with $L=15$. Reconstructed SH coefficients are obtained using the proposed multi-pass IRF, where the samples of the signal are taken as (a) 991 samples of the Equiangular sampling scheme, (b) 972 samples of the HEALpix sampling scheme, (c) $900$ random samples (d) 450 (e) 900 and (f) 1350 samples of the optimal dimensionality sampling scheme.}
\label{fig:sampling}
\end{figure*}

\section{Analysis of Multi-Pass IRF}

\subsection{Partition Choices}
In order to understand the partitioning of $\lsphL{L}$, we refer to the graphical representation of $\lsphL{L}$ shown in Fig. \ref{fig:spectral_domain_representation}, which describes the position of spherical harmonic coefficients with respect to degree $\ell \in (0,1,\hdots,L-1)$ and order $m \le |\ell|$.  We give numbers to the spectral harmonic coefficients (basis functions) shown in Fig. \ref{fig:spectral_domain_representation} from 1 to $L^2$ in a way that we start the domain from $\ell=0,m=0$ and then traverse the whole domain by $m=-\ell$ to $m=\ell$ for increasing values of $\ell$. In a similar way, we can also traverse the whole domain by fixing $m$ for all values of $\ell$.
We analyse four different type of partitions, whose sizes vary with the increasing or decreasing values of degrees $\ell$ and orders $m$. The size of each partition is denoted by $N_k$.  In all the partitions, the generalized IRF is run for all values of $k$ and for a fixed value of $i$.

\subsubsection*{Partition Choice 1}
We first consider the partitioning of $\lsphL{L}$ based on the spherical harmonic degree~\cite{Shen:2006}. We take $K=L$ partitions $\lsphL{L}^k$ each for degree $\ell=k-1$ such that the subspace $\lsphL{L}^k$ is spanned by spherical harmonics of degree $k-1$. Consequently, the dimension of each subspace is $N_k=2k-1$. As mentioned earlier, the IRF has been applied already for this choice of partition~\cite{Shen:2006}. We show through numerical experiments that alternative choices for partitioning result in faster convergence and more accurate computation of the SHT.
\subsubsection*{Partition Choice 2}
For partition choice 2, we combine the $k$-th partition choice 1  and $K-k+1$-th partition choice 1, to obtain $\frac{L}{2}$ or $\frac{L+1}{2}$ partitions for even or odd band-limit $L$ respectively. For even $L$, each partition 2 $\lsphL{L}^k$ is of size $N_k = 2L$ for $k=1,2,\hdots,\frac{L}{2}$. For odd $L$, we have $\frac{L+1}{2}$ partitions with $N_k = 2L$ for $k=1,2,\hdots, \frac{L-1}{2}$ and one partition of size $N_{\frac{L+1}{2}} L$.

\subsubsection*{Partition Choice 3}
Here, we consider partitioning with respect to each order $|m|<L$~(see \figref{fig:spectral_domain_representation}). Consequently, we have $2L-1$ partitions, one for each order $|m|<L$ and spanned by SH functions of order $m$.

\subsubsection*{Partition Choice 4}
Partition choice $4$ is obtained by combining the partitions in partition choice 3. We obtain $L$ partitions by combining partition choice 3 for $m$ and $-(L-m)$ for $m=1,2,\hdots L-1$. With such combining, we have $L$ partitions of $\lsphL{L}$ each of size $L$.


\subsection{Analysis}
Here we analyse the accuracy of the computation of the SHT, that is, the computation of SH coefficients, of the band-limited signal sampled over different sampling schemes. For the distribution of samples on sphere, we consider equiangular sampling~\cite{Equi:2011} and optimal-dimensionality sampling \cite{khalid:2014} in our analysis as these schemes support the accurate computation of the SHT for band-limited signals.
Among the sampling schemes on the sphere, which do not support the highly accurate computation of the SHT, we consider the HEALPix sampling scheme~\cite{Healpix:2005} and random samples with uniform distribution with respect to the differential measure $\sin\theta d\theta d\phi$.

In order to analyse accuracy, we take a test signal $g\in\lsphL{L}$ by first generating the spherical harmonic coefficients $\shc{g}{\ell}{m}$ with real and imaginary part uniformly distributed in $[-1,1]$ and using \eqref{Eq:f_expansion} to obtain the signal over the samples for each sampling scheme. For a meaningful comparison, we take approximately the same number of points for each sampling scheme. We apply the proposed multi-pass IRF for each choice of partition and each sampling scheme to compute the estimate of SH coefficients $\shc{\tilde g}{\ell}{m}$ and record the maximum error between reconstructed and original SH coefficients given by
\begin{align}\label{Eq:exp1:errors:max}
\epsilon_\textrm{max} &\dfn \max_{\ell<L, \, |m|\le\ell}        |\shc{g}{\ell}{m} - \shc{\tilde g}{\ell}{m} |,
\end{align}
which is plotted in logarithmic scale in \figref{fig:sampling} for band-limit $L=15$. Different partition choices and different sampling schemes~(see caption for number of samples for each sampling scheme) against the number of iterations of the proposed multi-pass IRF are plotted, where it can be observed that 1) the error converges to zero~($10^{-16}$, double precision) for all partition choices and sampling schemes, and 2) the error converges quickly for partition choice $4$. We also validate the formulation of the residual in \eqref{eq:res} by computing after each iteration of the multi-pass IRF. To illustrate the effect of the number of samples on the accuracy of the proposed multi-pass IRF, we have taken $2L^2$, $4L^2$ and $6L^2$ samples of optimal dimensionality sampling~\cite{khalid:2014} and plot the error $\epsilon_\textrm{max}$ in \figref{fig:sampling}(d)-(f), where it is evident that the error converges quickly for a greater number of samples. The convergence of the error is in agreement with the formulation of the residual in \eqref{eq:res}, however, convergence changes with the sampling scheme and nature of the partition of the subspace of band-limited signals. This requires further study and is the subject of future work.

%


\section{Conclusions}
We have presented the generalized iterative residual fitting~(IRF) method for the computation of the spherical harmonic transform~(SHT) of band-limited signals on the sphere. Proposed IRF is based on partitioning the subspace of band-limited signals into orthogonal spaces. In order to improve the accuracy of the transform, we have also presented a multi-pass IRF scheme and analysed it for different sampling schemes and for four different size partitions. We have performed  numerical experiments to show that accurate computation of the SHT is achieved by multi-pass IRF. For different partitions and different sampling distributions, we have analysed the residual~(error) and demonstrated the convergence of the residual to zero. Furthermore, it has been demonstrated that the rate of convergence of error depends on the sampling scheme and choice of partition. Rigorous analysis relating the nature of partitions and convergence of the proposed method and application of proposed method in medical imaging, computer graphics and beyond are subjects of future work.
	
	\renewcommand{\baselinestretch}{1}
	
	\bibliography{IRF_bib} 

\end{document}